# Optimization of Fuzzy Controller of a Wind Power Plant Based on the Swarm Intelligence

V.Z. Manusov[1],
P.V. Matrenin[1]
[1] Department of Industrial Power Supply System / Novosibirsk State Technical University,
Novosibirsk, Russia

*Abstract* – The article considers the problem of the optimal control of a wind power plant based on fuzzy control and automation of generating the fuzzy rule base. Fuzzy rules by experts do not always provide a maximum power output of the wind plant and fuzzy rule bases require an adjustment in the case of changing the parameters of the wind power plant or the environment. This research proposes the method for optimizing the fuzzy rules base compiled by various experts. The method is based on balancing weights of fuzzy rules into the base by the Particle Swarm Optimization algorithm. The experiment has shown that the proposed method allows forming the fuzzy rule base as an exemplary optimal base from a non-optimized set of fuzzy rules. The optimal fuzzy rule base has been taken under consideration for the concrete control loop of wind power plant and the concrete fuzzy model of the wind.

## I. INTRODUCTION

THE WIND POWER has been applied more often during the last decades. It is related to the fact that the wind is renewable and relatively ecologically-friendly power source. Cost of the power generated at the wind-driven power plants (hereinafter – WPP) shall be comparable to the cost of the power obtained from fossil fuels which have significantly decreased for the last 2 years. That is why it is quite important to increase the amount of the power entrapped at WPP for the purposes of the cost reduction.

The most important problems of WPP control are generation of maximum possible power in conditions of the wind rate variability and rotation speed limitation for the purposes of WPP mechanical damage prevention. The power $P$ generated by WPP at the wind velocity $u$, at the area covered by wind-wheel, $A$ and at the air density $\rho$ is determined by the air-flow power $E$ per unit time $t$ and can be expressed by the following equation [1]:

$$P = \frac{E}{t} = \frac{1}{2} C_P \frac{(Aut\rho)u^2}{t} = \frac{1}{2} C_P A\rho u^3$$

Where $C_P$ – power factor, determining flow energy utilization efficiency. Depending on the wind velocity and direction it is necessary to change WPP parameters, i.e. to perform its control aimed at $C_P$ value increasing and generated power $P$ maximization. For these purposes there are three main actions in WPP control loop [2]:
- turbine blades angle of attack;
- nacelle turning control;
- blades length changing.

Due to the fact that the wind velocity and direction always change, and considering that each of the control actions has its own peculiarities, it is necessary to construct controller with a help of which one can perform WPP automatic control to increase generated power amount [2, 3]. There are also various methods of optimal WPP control.

1. The method of peak power point tracking can be used in case of WPP parameters uncertainty, but its efficiency decreases as far as the gradients of the variables series in dynamic conditions are used [4].

2. Sliding control method, implementing non-linear control law, helps to save the required output variables effectively when external conditions change [2, 5]. Nevertheless, this method is not reliable as it may cause destructive mechanical vibrations of blades and other WPP elements [5].

3. Control with linear feedback. Although WPP is a non-linear system, it is possible to use the control with linearized feedback. Quite high computational complexity for real-time operation is the disadvantage of this method as far as linearization is performed by the high order polynomial [2].

4. Fuzzy control. For the purposes of WPP control, one shall take into consideration stochastic nature of the wind, non-linearity of the wind power transformation systems, availability of unknown parameters and WPP peculiarities [3]. The methods based on the fuzzy logic allow constructing of controllers which consider these factors. That is why the fuzzy controllers are more effective for WPP control [2]. But the problem of linguistic variables membership functions and fuzzy rule base formation arises.

## II. PROBLEM DEFINITION

Creation of the satisfactory functions of variables memberships for WPP is a simpler task than fuzzy rules formation. For the purposes of membership levels determination, it is not necessary to consider the variables interactions and perfect values searching is also not required, as it is the evaluation "fuzziness" that is used for the fuzzy control. It is necessary to involve experts in this subject domain to form fuzzy rule base. But the experts' evaluations may be in conflict because of subjectivity. Moreover, the best fuzzy rules set may vary for different WPP types as well as for different technical, geographical and climatic operation conditions.

Therefore, two related problems arise:
- selection from a variety of the rules compiled by several experts to create unified consistent rules base;
- adjustment of the existing fuzzy rule base to satisfy the requirements of the new WPP types or new conditions of usage.

Both these problems are confined to the problem of optimization of the fuzzy rules (called also **FR**) according to the specified optimality criterion and limitations. As a rule, the largest amount of the power generated at WPP within the considered time period is the criterion, and limitations are related to WPP operation stabilization. The problem of the best **FR** set searching can be solved with a help of mathematical models of the WPP under consideration and the wind for the purposes if optimal fuzzy controller formation designated for the certain WPPs and geographical location.

## III. THEORY

### A. Fuzzy model of the wind

The wind velocity is one of the most important parameters for the purposes of evaluation of the power which can be generated at WPP. To apply fuzzy logic in the wind-power engineering tasks first of all it is necessary to describe the wind velocity using the linguistic variables. The linguistic variables allow specifying of the variables values using words and sentences of the natural language. In the process of fuzzy models creation, one of the main stages is the construction of the membership function describing semantics of the variable base values.

It is known that the wind velocity can be represented by the Beaufort scale [2]. On this scale the wind velocity is divided into 9 intervals. The intervals for the weakest and the strongest winds can be excluded because there is no sufficient impact on the wind-driven power plant at the minimum wind velocity and the plants are switched off at the strong wind to prevent destruction. The Beaufort scale has to be described by the membership functions of the linguistic variables, i.e. it is necessary to specify what is the wind velocity $u$, what is the membership level and to which linguistic variable it is relative to. In the case under consideration so called LR-functions [6] which can be easily demonstrated in graphic form are used (fig. 1).

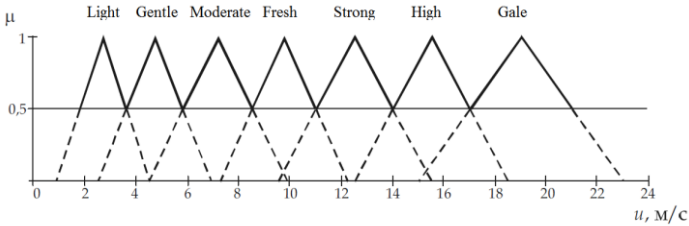

Fig. 1. Fuzzy wind velocities

Fig. 1 demonstrates that, for example, the wind velocity of 11 m/s can be related both to the fresh wind and to the strong wind, the wind velocity of 18 m/s is more likely to be related to the strong wind, but to some extent it can be related to the gale wind.

### B. Fuzzy controller of WPP power output

The fuzzy WPP controller is represented in fig. 2.

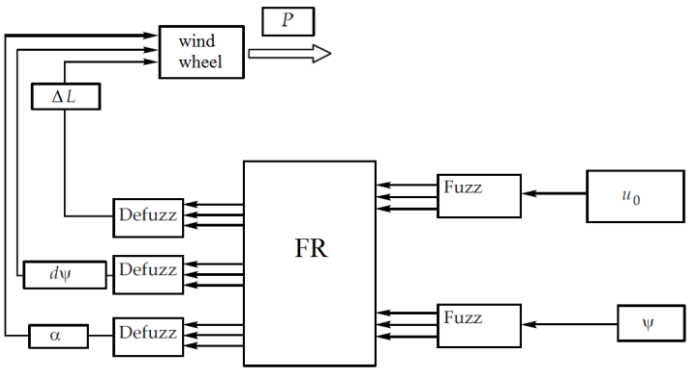

Fig. 2. Diagram of the fuzzy control of WPP power output

The wind velocity $u_0$ and direction $\psi$ (the angle between the wind direction and nacelle position) are fed at the controller input; the control commands for nacelle turning ($d_\psi$), blade angle of attack ($\alpha$) and windwheel blade length ($\Delta L$) are at the controller output. The membership functions of the wind velocity are shown in fig. 3, the membership functions of the wind direction are shown in fig. 4.

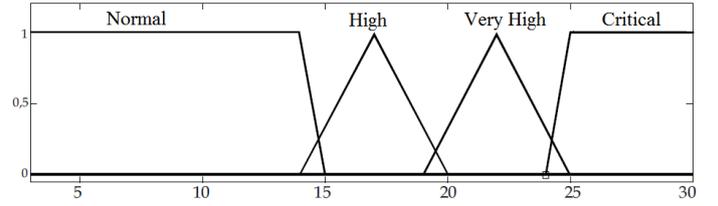

Fig. 3. The wind velocity membership functions

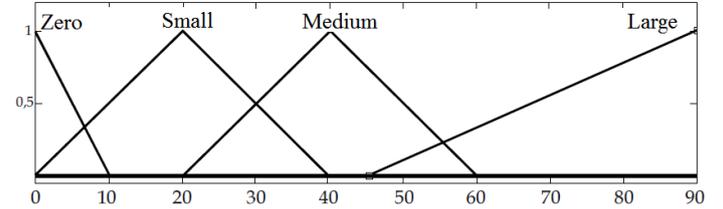

Fig. 4. The wind direction membership functions

To simplify further description let us identify the wind velocity from working to critical as $N$, $H$, $VH$, $Cr$, and angle $\psi$ from zero to large as $Z$, $S$, $M$, $L$. The membership functions for the output values are specified in a similar way: blade angle of attack (fig. 5, $Z$, $S$, $M$, $N$, $L$), blade length (fig. 6, $Z$, $S$, $M$, $L$) and nacelle turning (fig. 7, $NL$, $NS$, $Z$, $PS$, $PL$).

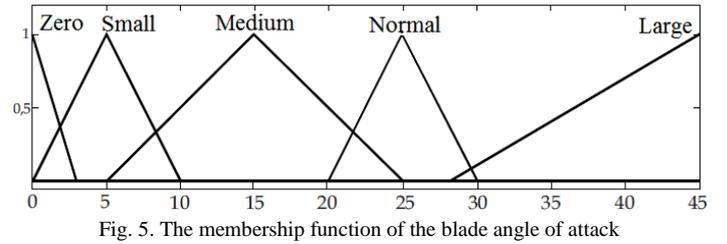

Fig. 5. The membership function of the blade angle of attack

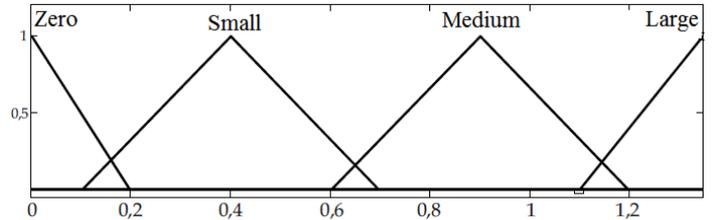

Fig. 6. The membership function of the blade length

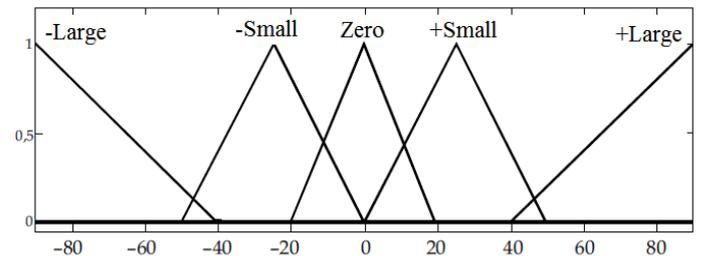

Fig. 7. The membership function of the nacelle turning

### C. Fuzzy controller operation process

Fuzzy controller operation is performed according to the following schedule (detailed description of the described controller operation is given in [2]).

1. The values of the wind velocity and direction pass the fuzzification stage, i.e. determination of the membership functions for the input variables. As a result of fuzzification the particular values of $u_0$ and $\psi$ are set according to the membership level. For example, if the angle $\psi$ has the value of 35 degrees it is small one with a membership of 0.25, middle one with a membership of 0.75 and zero one and large one with zero membership.

2. Clipping levels for every fuzzy rule prerequisite are selected; as a result, truncated membership functions for the output values of each rule are formed.

3. The determined truncated membership functions are joined together, as a result, some concluding fuzzy subsets are obtained for each output variable. The Mamdani algorithm is used for this purpose.

4. For every output value one shall perform defuzzification, i.e. determination of particular output value.

### D. Swarm Intelligence algorithms

The problem of the fuzzy rule base optimization is the high-dimensional optimization problem with non-differentiable objective function, set of limitations and internal interactions. Moreover, the problem is a stochastic one as far as one of the most important system element is the stochastic fuzzy wind model. For such problems the most effective are the population optimization methods based on the agents systems usage. In this particular case, an agent is meant some point in the problem solution searching space, and the optimization process consists in the agents' transposition within this space.

The population algorithms can be divided into evolutionary ones and swarm ones. Both swarm intelligence and evolutionary algorithms belong to population stochastic algorithms and can be described using Markov's finite chains. The swarm algorithms use collective decentralized transposition of the agents of the same population without selection, destruction and creation of new spawns. Any swarm intelligence algorithm can be represented as [8]

$$SI = \{S, M, A, P, I, O\} \quad (1)$$

where **S** – set of swarm agents;
**M** – material for experience interchange among **S** agents;
**A** – swarm algorithm operation rules;
**P** – parameters used in **A** rules;
**I** and **O** – swarm algorithm input and output, by means of which it interacts with the problem being solved and the control system.

Swarm intelligence includes such algorithms as:
- Gravitational Search algorithm;
- Ant Colony Optimization algorithms;
- Bat algorithm;
- Bacteria Swarm Optimization algorithm;
- Artificial Bee Colony Optimization algorithm;
- Particle Swarm Optimization algorithm;
- Fish Shoal Search;
- Electromagnetic Search.

The present work uses particle swarm optimization algorithm (PSO [7]), as it is one of the best recommended in practice in various fields.

### E. Basic principles of the particle swarm optimization algorithm

In 1995 James Kennedy and Russel Eberhart proposed the method based on the birds' flock behavior imitation for the optimization problems approximate solution [7]. The flock behavior is well-coordinated, every bird follows some rules. If a bird finds some food it announces it to the whole flock and the flock directs to that place. Due to the data exchange between each bird and the whole flock, a kind of the flock collective behavior which is called "swarm intelligence" is formed. It is necessary to understand that it is just the simplified model of the birds' behavior, which can be incorrect from the ornithological point of view, nevertheless, it is acceptable in practice for the optimization problems solving in various fields.

### F. Mathematical model of the particle swarm optimization algorithm

When passing to the algorithm mathematical model the word "bird" is replaced by the word "agent" to provide more formal description and the following rules are introduced:
- at a certain moment each agent is located at some point within the space which is specified by **X** coordinate vector;
- each agent is characterized by **V** velocity vector;
- agents assess their location within the space by the value of a criterion $f(\mathbf{X})$;
- each agent knows its position within the space **PX**, in which it was located and in which the value $f(\mathbf{X})$ was maximum (the best position);
- each agent knows the best position of all the agents within the space **PX** (**GX** position in which the value $f(\mathbf{X})$ is maximum)
- agents tend to the best **PX** positions, in which they had been located and to the best general position **GX**;
- stochastic factors and inertia influence the agents' velocity.

Let us suppose that the problem of $f(\mathbf{X})$ function maximum searching is being solved. In this function **X** is the vector of variate parameters, which can take the values from some **D** area. In PSO algorithm according to the expression (1) **S** = {set of agents}, **M** = **GX**, **A** = algorithm operation rules given below, **P** = algorithm parameters, which are also given below ($\alpha_1$, $\alpha_2$, $\omega$, $v_{max}$), **I** = {$f(\mathbf{X})$, **D**}, O = {**GX**, $f(\mathbf{GX})$}.

PSO algorithm is terminated when the specified number of iterations is achieved or when a satisfactory solution is reached, or in such case when no improvement of already found a solution cannot be found within a certain time period. So the algorithm can be written down the following way.

1. Iteration number equals one. Allocate the agents accidently within the solution area **D**, define original velocities (can be zero or accident). **PX** positions of each agent equal **X**, **GX** position equals **PX** position of any agent.

2. Calculate the values of the optimized function $f(\mathbf{X})$ for every agent. If the determined value is more than $f(\mathbf{PX})$ or $f(\mathbf{GX})$, the values **PX** and **GX** shall be updated correspondingly.

3. Calculate new values of **V** velocity for every agent with regard to the limitation factor $v_{max}$.

4. Calculate new **X** coordinates for every agent.

5. If the termination condition has not been fulfilled, finish the algorithm, otherwise – go to the step 2 and increase iterations number.

The position **GX** is a result of the algorithm performance. The steps 3 and 4, namely the rules (formulas) for velocities and positions calculation, are the most interesting. There are a great number of these formulas variations, let us demonstrate one of the simplest. If the values **X**, **V**, **PX** are known for some agent and **GX** is known for the whole swarm, the new velocity value for each agent at step 3 can be determined the following way:

$$\mathbf{V} = \omega\mathbf{V} + \mathbf{rnd}_1(\mathbf{PX} - \mathbf{X})\alpha_1 + \mathbf{rnd}_2(\mathbf{GX} - \mathbf{X})\alpha_2 \quad (2)$$

where $\alpha_1$, $\alpha_2$, $\omega$ – algorithm parameters, determining the swarm behavior peculiarities, $\mathbf{rnd}_1$ and $\mathbf{rnd}_2$ – the vectors of random real numbers, the elements of which are allocated uniformly from 0 to 1, their dimensions equal dimensions of the solution search-space. At the same time, the limitations for the maximum module velocity ($v_{max}$) are determined [9].

New agent position on step 4 will be determined as:

$$\mathbf{X} = \mathbf{X} + \mathbf{V} \quad (3)$$

If **X** is outside the solution search-space borders, the nearest acceptable value is usually assigned for **X**. Detailed PSO description is given in [8].

### G. Rule base formation

The traditional approach for formation of the fuzzy rule base determining controlled object behavior supposes formalization of the expert's experience or experts' group experience in the considered domain. As a result, the set of fuzzy rules is formed. Among these rules is: "IF the wind velocity is strong and the nacelle turning angle is small, THEN blade angle of attack is middle one". But every expert put his internal representation of the system and of the fuzzy variables values. Moreover, the experts have different competency, as a result joining of all expert evaluations to the unified rule base is a sophisticated problem and, as it has been mentioned before, the related problem of already created base adaptation to new WPPs, operation mode changes or new geographical region for WPP designing arises.

Let us suppose that there are some mathematical model of the considered control object and the wind direction and velocity observation results, so one can verify the controller quality by means of feeding of the mentioned observation results to the model input and receive the output the optimality criterion, such as the power generated at WPP according to observation samples within quite a long period, for example, 1 year. In such case, any fuzzy rule base **FR** can be evaluated with the value of criterion $f(\mathbf{FR})$. As a result, the method of the fuzzy rule base evaluation is obtained but the problem of their formation remains. This problem can be automated partially or fully with the help of artificial intelligence methods, particularly – swarm intelligence.

Full automation supposes creation of the fuzzy rule base without the experts' involving. If the rule base **FR** is encoded with a help of the vectors of integral or real numbers **X**, it is possible to calculate the optimality criterion $f(\mathbf{FR}) = f(\mathbf{X})$, the value of which shall be maximized. As a result, the problem of continuous-discrete optimization arises. The problem can be solved with PSO algorithm. It is necessary to point out that the swarm intelligence algorithms do not depend on the internal logic of the optimality criterion calculation. This flexibility is one of the main advantages of the swarm intelligence [7, 9]. But the problem of the fuzzy rule base coding and decoding by **X** numbers vector takes considerable time and labor. A detailed description of this problem solving method can be made in the separate work. That is why we consider below the simpler optimization method of the existing fuzzy rule base which may contain conflicts and faults and can even be generated randomly.

## IV. EXPERIMENTAL RESULTS

### A. Usage of the particle swarm optimization algorithm for the fuzzy rule vase arrangement

Let us assume that every expert creates the fuzzy rule base independently or with a help of the fuzzy logic specialist. At the same time, as far as the fuzzy rule base will be subsequently improved with a help of the swarm intelligence, the experts are instructed to act in a free manner and to propose the rules in which they are not certain. As a result, the fuzzy rule base contains such rules which are not included when the traditional approach of their formation is used. Then all the rules are unified into a single fuzzy rule base and significance factor equal one is assigned to every rule. The more is the rule weight, the bigger is its influence on the taken managerial decision. If an amount of all the rules equals $m$, then their weight vector with dimensionality $m$ will be obtained. One can change the rules weight in **FR** base and, consequently, the base quality $f(\mathbf{FR})$, by means of the vector elements changing,

If the weight vector is indicated by **X**, then the maximization problem $f(\mathbf{X})$ is obtained. This problem can also be solved by precise methods but only when its dimensionality is not big because the time required for this problem solving increases exponentially as far as its dimensionality increases as it belongs to NP-hard problems class. For example, for only 20 rules at possible weight values from 0 to 9 with step 1 the number of possible solutions amounts $20^{10}$; when the time for one rule base evaluation is one microsecond, 119 days is required for complete searching. Use of directed searching methods can significantly reduce solution time period but is still is quite large-scale and moreover significant time and labor costs are required for these methods adaptation to the considered sophisticated problem. That is why it is reasonable to apply the swarm intelligence algorithms allowing finding of the nearest to the best solution within an acceptable time period (depending on the problem difficulty and dimensionality as well as on computational capability from few seconds to several hours).

To apply PSO algorithm it is necessary only to determine the way how **X** vector will determine rules weight. Rule weight is the value describing the influence of the rule on the output membership functions determination, i.e. when the center of mass is determined by the Mamdani algorithm. In other words, when passing to geometrical meaning, the weight of each rule is a measure of its density. Let the rules weights to be taken in real values from 0 to 1. In such case the more is the weight the more is the rule influence on the final result, which means that after optimization with PSO algorithm the most representative models of WPP and the wind will have the biggest influence on the created fuzzy controller and consequently on WPP performance. Moreover, it is reasonable to introduce a minimum limit for the rule weight $b$. If the rule weight is lower than this limit it is just not used. In result of PSO algorithm operation, the rules weight vector will be obtained after which the rules having the weight lower than $b$ ($\mathbf{X}_i < b$) will be deleted from **FR** base.

As a result of cutoff by the limit, the rules in conflict and the rules having bad influence on the power generated by WPP will be deleted.

### B. Formal description of the proposed method

The following pseudo-code gives the formal description of the developed method.

1. Input and generation of the original fuzzy rule base $\mathbf{FR}_0$.
2. Initialization of the agents set **S** of PSO algorithm.
3. The cycle before execution of the specified number of PSO algorithm iterations.
3.1. The cycle for all **S** agents.
3.1.2. Create fuzzy rules weights vector: **W** = **X**.
3.1.3. Cutoff the rules from $\mathbf{FR}_0$, with weight $\mathbf{W}_i$ lower than $b$.
3.1.4. Calculate the quality criterion for the obtained rule base $f(\mathbf{FR}_0)$.
3.1.5. Update **PX** and **GX** vectors, if $f(\mathbf{FR}_0)$ is bigger than the criterion value in the agent position **PX** of the whole swarm position **GX**, correspondingly.
3.1.6. Return to the rule base all the rules which have been cutoff on the step 3.1.3.
3.2. The cycle for all **S** agents.
3.2.1. Update the agent velocity **V** by the formula (2).
3.2.2. Update the agent position **X** by the formula (3).

4. Form the final fuzzy rule base.
4.1.2. Create fuzzy rules weights vector: **W** = **GX**.
4.1.3. Cutoff the rules from **FR**$_0$, with weight **W**$_i$ lower than *b*.
4.1.4. Add the remaining rules to the final base **FR**$_1$.
4.1.4. Calculate the quality criterion for the obtained rule base *f*(**FR**$_1$).

Therefore, on the algorithm input we have non-optimized fuzzy rule base which may contain conflicts and faults and on the output we have the fuzzy rule base optimized by the specified criterion, WPP models, and the wind models.

*C. Usage of the model fuzzy rule base*

At the present working stage, the complete mathematical model of WPP is not developed yet, but in the previous works such as [2], the fuzzy rule base for WPP control according to the diagram given in fig. 1 has already been formed. This rule base has been developed by expertise and it may be considered as the model base. That is why for the purposes of experimental evaluation of the developed method applicability they have set the problem of the method application for the automatic formation of the rule base as close to the model **FR\*** as possible from the non-optimized fuzzy rule base **FR**$_0$. The base **FR**$_0$ is developed by means of the random new rules adding to the base **FR**$^*$. To compare the rule bases it is better to pass on the weight vectors at the same time their dimensionalities shall coincide. For this end the model rule base can also be represented by the weights vector **W**$^*$, for which the weights of existing rules equal one and of all the others equal zero. In such case the rule base evaluation criterion can be written the following day:

$$f(\mathbf{W}) = \sum_{i=1}^{m} |W_i - W_i^*| \quad (4)$$

The expression (3) *m* reflects a number of the rules in non-optimized base **FR**$_0$. It is necessary to note that before *f*(**W**) calculating, the weight having values below the limit shall be zeroed which corresponds to deleting of the corresponding rule from the base. It also means that in the specified formulation the problem consists in minimization of the rule base deviation from the model base.

The conducted experiments have used the fuzzy rule base [2] shown in the Table I and written in format: "if the wind velocity $u_0$ equals the value indicated in the column and the wind direction $\psi$ equals the value indicated in the line, then the required values of the angle of attack, blade length change, and the nacelle angle and turning are indicated in the cell on the intersection of this column and this line".

TABLE I
MODEL FUZZY RULES BASE

| $\psi$ | $u_0$ | | | |
|---|---|---|---|---|
| | N | H | VH | Cr |
| Z | a = Z<br>ΔL = L<br>dψ = Z | a = M<br>ΔL = Z<br>dψ = Z | a = L<br>ΔL = Z<br>dψ = Z | a = L<br>ΔL = Z<br>dψ = PL |
| S | a = Z<br>ΔL = L<br>dψ = Z | a = S<br>ΔL = Z<br>dψ = Z | a = M<br>ΔL = Z<br>dψ = Z | a = L<br>ΔL = Z<br>dψ = PL |
| M | a = Z<br>ΔL = L<br>dψ = NS | a = S<br>ΔL = Z<br>dψ = Z | a = M<br>ΔL = Z<br>dψ = Z | a = L<br>ΔL = Z<br>dψ = PS |
| L | a = L<br>ΔL = L<br>dψ = NL | a = Z<br>ΔL = Z<br>dψ = Z | a = S<br>ΔL = Z<br>dψ = Z | a = L<br>ΔL = Z<br>dψ = Z |

Viewing the columns and the lines of the Table I the following fuzzy rules definitions will be obtained:

1) if the wind velocity *N* (Normal) and direction *Z* (Zero), then the angle of attack *α* shall be *Z* (Zero), blade length Δ*L* shall be *L* (Large) and nacelle turning *dψ* shall be *Z* (Zero);
2) if the wind velocity *N* and direction *S*, then the angle of attack shall be *Z*, blade length - *L* and nacelle turning - *Z*;
…
16) if the wind velocity *Cr* and direction *L*, then the angle of attack shall be *L*, blade length - *Z* and nacelle turning - *Z*.

There is the total of 16 rules in the model base. The non-optimized base contains 184 more rules so the total number (*m* in the expression (4) equals 200.

PSO algorithm with 50 agents, 1000 iterations has been used in the experiment. Algorithm parameters were $α_1$ = 1.5, $α_2$ = 1.5, $ω$ = 0.729, $v_{max}$ = 0.1, the optimization process diagram (gradual solution improvement, i.e. the original fuzzy rule base approximation to the model base) is given in fig. 8.

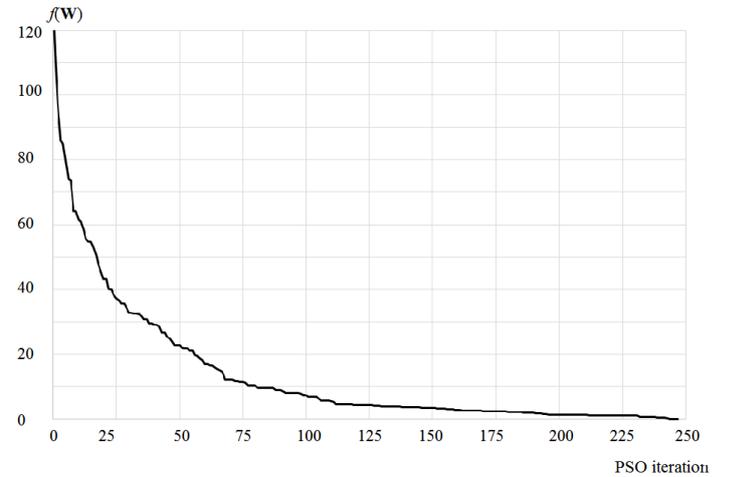

Fig. 8. The optimization process of the rule base **FR**$_0$

V. DISCUSSION OF RESULTS

The fig. 8 demonstrates that after 250 iterations of PSO algorithm operation the original fuzzy rule base deviation from the model base equals 0, i.e. the required rule base was obtained automatically from the non-optimized base consisting of 200 rules. Due to stochastic nature of PSO algorithm, it can produce different results from start to start. That is why 10 more experiments have been conducted. In 9 of the experiments the algorithm has successfully optimized the original rule base for absolute matching with the model base within the time limits. In the remaining experiment the final base has the number of rules exceeding the number of rules in the model base only by 2 rules. At the same time a small number of PSO agents (50) and a small number of iterations (less than 1000) allows finding the problem solution with a help of a personal computer in few seconds depending on its computational capability, and on the processor Core-i7 (2.4 GHz) the problem solving takes 1-2 seconds.

It is necessary to note that the experiments have demonstrated the need for significant limitation of the maximum PSO agents' velocity ($v_{max}$) in the process of this problem solving, because when the value $v_{max}$ exceeds 0.5 the optimal solution has not always been found even at 100 thousand of PSO iterations.

## VI. CONCLUSION

The article considers the problem of optimal WPP control on the basis of a fuzzy controller from the point of view of the fuzzy rule base automatic formation. The rules of the base provide maximization of the power generated at WPP and fuzzy controller adaptation to the parameters of WPP and external environment.

The optimization method of the fuzzy rule base compiled by different experts is proposed to cutoff contradictory and wrong fuzzy rules and significance weights distribution among remaining fuzzy rules to provide WPP control efficiency improvement. The method is based on the presentation of the considered problem as the discrete-continuous problem of fuzzy rules weight optimization. The problem discreteness is related to deleting from the base of the rules weight of which is lower than the certain limit.

The particle swarm optimization algorithm is used to solve this problem. This algorithm allows forming of the base from the base consisting of 200 rules which coincides with the base of 16 fuzzy rules selected as the model base in the process of experiments conducted.

It may be concluded that the problem of quality fuzzy rule base development can be significantly simplified via automation by means of artificial intelligence particularly by means of the particle swarm optimization algorithm.

In future, it is proposed to verify the proposed approach on WPP mathematical models to obtain the fuzzy rule base providing the largest mathematical expectation of the generated power.

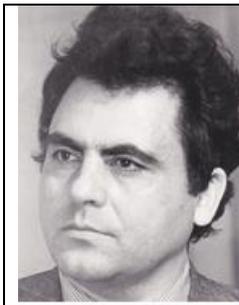

Vadim Zinovievich Manusov received the B.S. and the PhD degrees electrical engineering from Novosibirsk Electric Technical Institute, Novosibirsk, Russia in 1963 and 1986, respectively.

He is a Professor of the Department of Power Supply Systems in Novosibirsk State Technical University, Russia.

His current research area is artificial intelligence technologies and probabilistic methods in electric power systems.

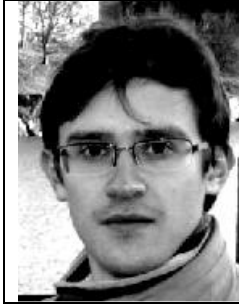

Pavel Viktorovich Matrenin received the B.S. and M.S. degrees information technologies from Novosibirsk State Technical University, Novosibirsk, Russia in 2012 and 2014, respectively. He is a PhD student in Novosibirsk State Technical University.

His main interests are related to stochastic optimization methods, design and development information systems, and artificial intelligence technologies in electric power systems.